\documentclass[twocolumn,english,aps,arxiv,superscriptaddress]{revtex4-2}
\usepackage{grffile}
\usepackage[latin9]{inputenc}
\usepackage{color}
\usepackage{babel}
\usepackage{graphicx}
\usepackage{epstopdf}
\usepackage{amsmath}
\usepackage{cancel}
\newcommand{\bet}{\begin{table}[hbt]\centering}
\PassOptionsToPackage{normalem}{ulem}
\usepackage{ulem}
\usepackage[unicode=true, bookmarks=false, breaklinks=false,pdfborder={0 0 1},colorlinks=false] {hyperref}
\usepackage{breakurl}

\begin{document}

\title{Unusual dynamic susceptibility arising from soft ferromagnetic domains in MnBi$_8$Te$_{13}$ and Sb-doped MnBi$_{2n}$Te$_{3n+1}$ $(n=2, 3)$ 
 }

\author{Chaowei Hu}
\affiliation{Department of Physics and Astronomy and California NanoSystems Institute, University of California, Los Angeles,
CA 90095, USA}

\author{Makariy A.~Tanatar}
\affiliation{Ames Laboratory US DOE, Ames, Iowa 50011, USA}
\affiliation{Department of Physics and Astronomy, Iowa State University, Ames, Iowa 50011, USA}

\author{Ruslan Prozorov}
\affiliation{Ames Laboratory US DOE, Ames, Iowa 50011, USA}
\affiliation{Department of Physics and Astronomy, Iowa State University, Ames, Iowa 50011, USA}

\author{Ni Ni}
\email{Corresponding author: nini@physics.ucla.edu}
\affiliation{Department of Physics and Astronomy and California NanoSystems Institute, University of California, Los Angeles,
CA 90095, USA}

\begin{abstract}

MnBi$_{2n}$Te$_{3n+1}$ (MBT) is the first intrinsic magnetic topological insulator and is promising to host emergent phenomena such as quantum anomalous Hall effect. They can be made ferromagnetic by having $n\ge4$ or with Sb doping. We studied the magnetic dynamics in a few selected ferromagnetic (FM) MBT compounds, including MnBi$_8$Te$_{13}$ and Sb doped MnBi$_{2n}$Te$_{3n+1}$ ($n=2, 3)$ using AC susceptibility and magneto-optical imaging. Slow relaxation behavior is observed in all three compounds, suggesting its universality among FM MBT. We attribute the origin of the relaxation behavior to the irreversible domain movements since they only appear below the saturation fields when ferromagnetic domains form. The very soft ferromagnetic domain nature is revealed by the low-field fine-structured domains and high-field sea-urchin-shaped remanent-state domains imaged via our magneto-optical measurements. Finally, we ascribe the rare ``double-peak" behavior observed in the AC susceptibility under small DC bias fields to the very soft ferromagnetic domain formations. 

\end{abstract}
\date{\today}
\maketitle

\section{Introduction}

Magnetic topological insulators provide a fruitful playground for the interplay between magnetism and band topology, hosting emergent topological state such as axion insulators, magnetic Weyl semimetals and
Chern insulators\cite{tokura2019magnetic,he2018topological,liu2016quantum,wang2015quantized}. MnBi$_{2}$Te$_{4}$ was found to be the first intrinsic antiferromagnetic topological insulator with layered van der Waals (vdW) structure\cite{lee2013crystal, rienks2019large,zhang2019topological,li2019intrinsic,
otrokov2019prediction,
aliev2019novel,deng2020high-temperature,gong2019experimental,lee2019spin,
yan2019crystal,zeugner2019chemical,otrokov2019unique,zhang2019topological,chen2019intrinsic,
hao2019gapless,ge2020high,deng2020quantum,liu2020robust,li2020competing,ding2020crystal}. It has an A-type antiferromagnetic (AFM) structure. Within each MnBi$_{2}$Te$_{4}$ septuple layer (SL) the Mn moments align ferromagnetically (FM) while the SLs are antiferromagnetically aligned. It is, however, preferable to make the interlayer coupling to be FM so that dissipationless quantum anomalous Hall effect (QAHE) can be realized at zero field. To reduce the interlayer AFM coupling, one strategy is to add nonmagnetic Bi$_2$Te$_3$ quintuple layers (QL) between SLs to physically separate the magnetic MnBi$_{2}$Te$_{4}$ layers to get MnBi$_{2n}$Te$_{3n+1}$ (MBT). By this design, MnBi$_4$Te$_7$ and MnBi$_6$Te$_{10}$ are found to be weakly-coupled A-type AFM\cite{147,wu2019natural,ding2020crystal,
shi2019magnetic,tian2019magnetic,yan2020type,gordon2019strongly,hu2020universal,xu2019persistent,jo2020intrinsic,
tian2019magnetic,klimovskikh2020tunable}, and ferromagnetism is achieved when $n\ge 4$ for MnBi$_{8}$Te$_{13}$ and higher $n$ members \cite{1813,ding2021neutron}. The other strategy is to dope Bi by Sb. In practice, Mn$_{\rm{(Bi, Sb)}}$ antisites are produced and enhanced with the Sb doping due to the tendency of Sb and Mn to form antisites. The presence of antisites introduces additional Mn sublattices which facilitates FM coupling of the dominant Mn sublattice \cite{liu2020site,wimmer2020ferromagnetic,ge2021direct, Sb-Mn147}. As a result, the magnetic transition of the dominant Mn sublattice becomes FM-like in high-antisite MnSb$_2$Te$_4$ \cite{liu2020site,wimmer2020ferromagnetic,ge2021direct}, Sb-doped Mn(Bi$_{1-x}$Sb$_{x}$)$_4$Te$_{7}$ \cite{Sb-Mn147}, and Sb-doped Mn(Bi$_{1-x}$Sb$_{x}$)$_6$Te$_{10}$ \cite {wu2020toward}. 

Despite the tremendous efforts in studying magnetic properties of the MBT family, most works have been only about the steady-state measurement. For practical applications, especially in the pursuit of high-temperature QAHE when fluctuations become important, the study on magnetic dynamics is indispensable. Previously, a slow magnetic relaxation behavior was found in Mn(Bi$_{0.7}$Sb$_{0.3}$)$_6$Te$_{10}$ at low temperatures. This behavior is attributed to the vanishing interlayer coupling due to the large interlayer distance between SLs for $n\ge2$ \cite{wu2020toward}. This so-called single-layer magnetism picture is a two-dimensional analog of the single-molecule magnet, a typical system to exhibit the superparamagnetic (SPM) behavior. In the other limit, the slow relaxation dynamics is also found in FM MnSb$_2$Te$_4$ \cite{li2021spin}, where the interlayer distance is much smaller. A frequency-dependent peak shift $\chi'(T)$ is attributed to a spin glass (SG) state, but the mechanism of the glassiness is unclear. 

To understand the origin of the relaxation behavior in this family, we perform a comprehensive study of the dynamical magnetic properties in FM MnBi$_8$Te$_{13}$, Mn(Bi$_{0.93}$Sb$_{0.07}$)$_6$Te$_{10}$ and Mn(Bi$_{0.24}$Sb$_{0.76}$)$_4$Te$_7$ by the AC susceptibility and magneto-optical image measurements. We show that the slow relaxation behavior is universal among all FM MBT members, which arises from the dynamics of irreversible domain wall movements rather than the SPM or SG scenarios. Our study suggests FM domains in MBTs are very soft and weakly pinned, resulting a unique ``double-peak" behavior in the real part of the AC susceptibility $\chi'(T)$ under DC fields. 

\section{Experimental Methods}

Single crystals were synthesized using the self-flux method \cite{1813,Sb-Mn147}. The doping level was determined by the Wavelength-Dispersive-Spectroscopy (WDS) performed on a JEOL JXA-8200 Superprobe. DC and AC magnetization data were measured in a Quantum Design Magnetic Properties Measurement System (QD MPMS). AC susceptibility under DC field bias was performed after the system is field-cooled under the  corresponding DC bias field.

Direct magneto-optical Kerr imaging was performed using a helium flow-type
cryostat with temperature down to 5 K and the ability to apply a 1
kOe - range magnetic field. Optical imaging in linearly polarized
light was done using a Leica DMLM polarized microscope equipped with
the high quality polarizer and analyzer.
In the experiment, the sample is positioned on top of a gold-plated
copper stage with its flat surface perpendicular to the direction
of light propagation. The light polarization direction, also perpendicular
to the light propagation direction (thus parallel to the sample surface
that is being imaged), was controlled by a polarizer and could be
changed with respect to the stationary sample for optimal contrast.
Upon reflection off the sample, polarization direction rotates by
the angle proportional to the surface magnetization (magneto-optical
polar Kerr effect). Due to chirality of the problem, opposite magnetic
moments lead to the opposite directions of the polarization rotation.
When viewed through an analyzer rotated almost perpendicular to the
polarizer, a 2D image of the magnetic pattern emerges. The appearance
can be switched at will by adjusting the polarizer/analyzer pair and
we chose to show domains along the direction of the applied field
to be dark, whereas opposite domains are white and not magnetized
state is neutral-gray. In this setting, the maximum contrast between
the domains is achieved. More detailed discussion of magneto-optical
techniques can be found elsewhere \citep{Hubert2008}.

\section{Experimental Results}
\subsection{Magnetic relaxation revealed by AC susceptibility}
\begin{figure}
\centering
\includegraphics[width=3.4in]{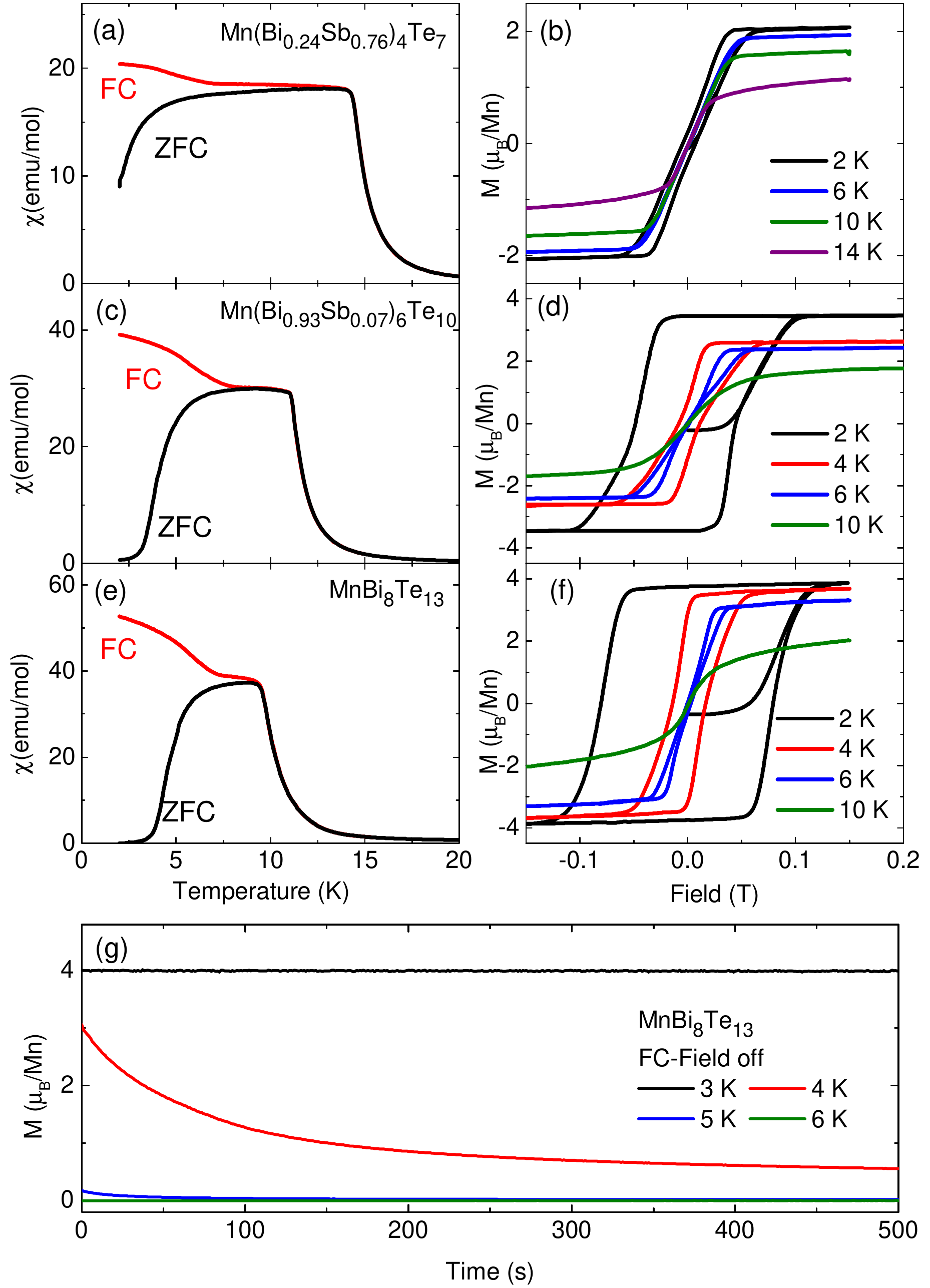} \caption{(a)(c)(e) The
temperature-dependent zero-field-cooled (ZFC) and field-cooled (FC) magnetic
susceptibility under 0.01 T with $H // c$ ; (b)(d)(f) The isothermal magnetization $M(H)$ at various temperatures. (g) The time dependence of the magnetization at various temperatures after MnBi$_8$Te$_{13}$ is FC under 0.1 T and then switched off. }
\label{fig:dc-xt}
\end{figure}

\begin{figure*}
\centering
\includegraphics[width=6.8in]{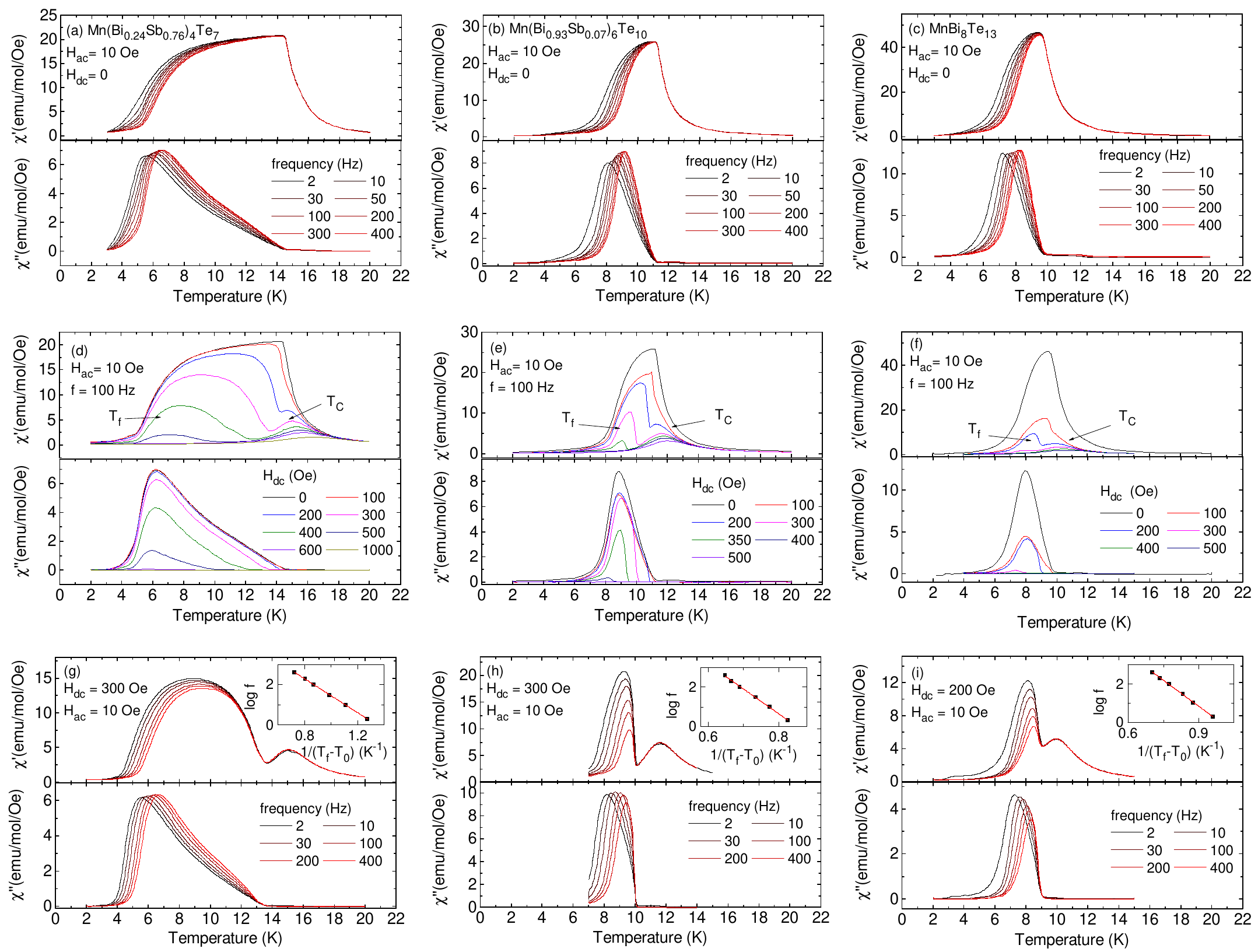} \caption{(a)-(c) The
temperature-dependent AC susceptibility for Mn(Bi$_{0.24}$Sb$_{0.76}$)$_4$Te$_7$, Mn(Bi$_{0.93}$Sb$_{0.07}$)$_6$Te$_{10}$ and MnBi$_8$Te$_{13}$ with $H // c$ at different sweeping frequency at zero DC field. (d)-(f) The
temperature-dependent AC susceptibility for the three compounds measured under fixed frequency and different DC field bias. (g)-(i) The
temperature-dependent AC susceptibility for the three compounds at fixed DC field bias and AC field, but with different sweeping frequency. Insets: Vogel-Fulcher fitting showing linearity between the log of $f$ and 1/(T$_f$-T$_0$). }
\label{fig:ac-xt}
\end{figure*}

In Figure 1, we show the zero-field-cooled (ZFC) and field-cooled (FC) DC temperature-dependent susceptibility $\chi(T)$ and isothermal magnetizations $M(H)$ of the three compounds. The $T_C$ determined at the sharp turn in $\chi(T)$ \cite{1813,Sb-Mn147} are at 14.2 K, 11.1 K and 10 K respectively. The difference in the ordering temperatures among the three, from Mn(Bi$_{0.24}$Sb$_{0.76}$)$_4$Te$_7$ to MnBi$_8$Te$_{13}$, lies in both the increasing interlayer distance and the decreasing extent of the Mn$_{\rm{(Bi, Sb)}}$ antisites. The larger the distance, the weaker the interlayer interaction, so $T_C$ is the lowest in MnBi$_8$Te$_{13}$. As antisites switch the dominant Mn sublattice from AFM to FM in MnBi$_4$Te$_{7}$ and MnBi$_6$Te$_{10}$, they also increase the overall interlayer interaction and hence give rise to the higher $T_C$ than the N$\acute{\rm e}$el temperature of the parent compounds \cite{147}. In the $M(H)$ data presented in Figs. 1 (b)(d)(f), FM hysteresis can be seen at 2 K for all compounds, and is the smallest in Mn(Bi$_{0.24}$Sb$_{0.76}$)$_4$Te$_7$ and the largest in MnBi$_8$Te$_{13}$. The softness of a FM is controlled by two factors, one is the magnetocrystalline anisotropy, the other is defect. The crystalline anisotropy is proportional to the coercive field according to the Stoner-Wohlfarth model for a single-domain FM, while the defects tend to provide more pinning. Thus the higher the magnetocrystalline anisotropy, the harder the FM; the more the defects, the harder the FM. Since the saturation field of MnBi$_{8}$Te$_{13}$ with $H	// ab$ is about 3 times of that for Mn(Bi$_{0.24}$Sb$_{0.76}$)$_4$Te$_7$, the former has higher magnetocrystalline anisotropy \cite{Sb-Mn147,1813}. Meanwhile, neutron scattering data suggest that Mn(Bi$_{0.24}$Sb$_{0.76}$)$_4$Te$_7$ has more defects than MnBi$_{8}$Te$_{13}$ \cite{Sb-Mn147,1813}. Therefore, the fact that MnBi$_{8}$Te$_{13}$ is the hardest among all three suggests that the magnetocrystalline anisotropy dominates the level of softness in MBT, which indicates that the domain pinning caused by the defects is very weak in MBT. We also note that the hysteresis curves are non-trivial with the ``bow-tie"-shaped hysteresis most clearly demonstrated at around 6 K in Fig. 1(d) and (f). This unusual hysteresis loop can be linked to the anomalous behavior of the FC $\chi(T)$. Instead of a smooth increase with establishing FM order upon cooling, FC $\chi(T)$ first shows a flat plateau below $T_C$ before a clear slope change can be observed in the FC $\chi(T)$ near 6 K. These anomalous observations are likely to be caused by possible non-trivial domain formation in MBT, as will be discussed later. 

\begin{figure*}[ptb]
\centering
\includegraphics[width=7in]{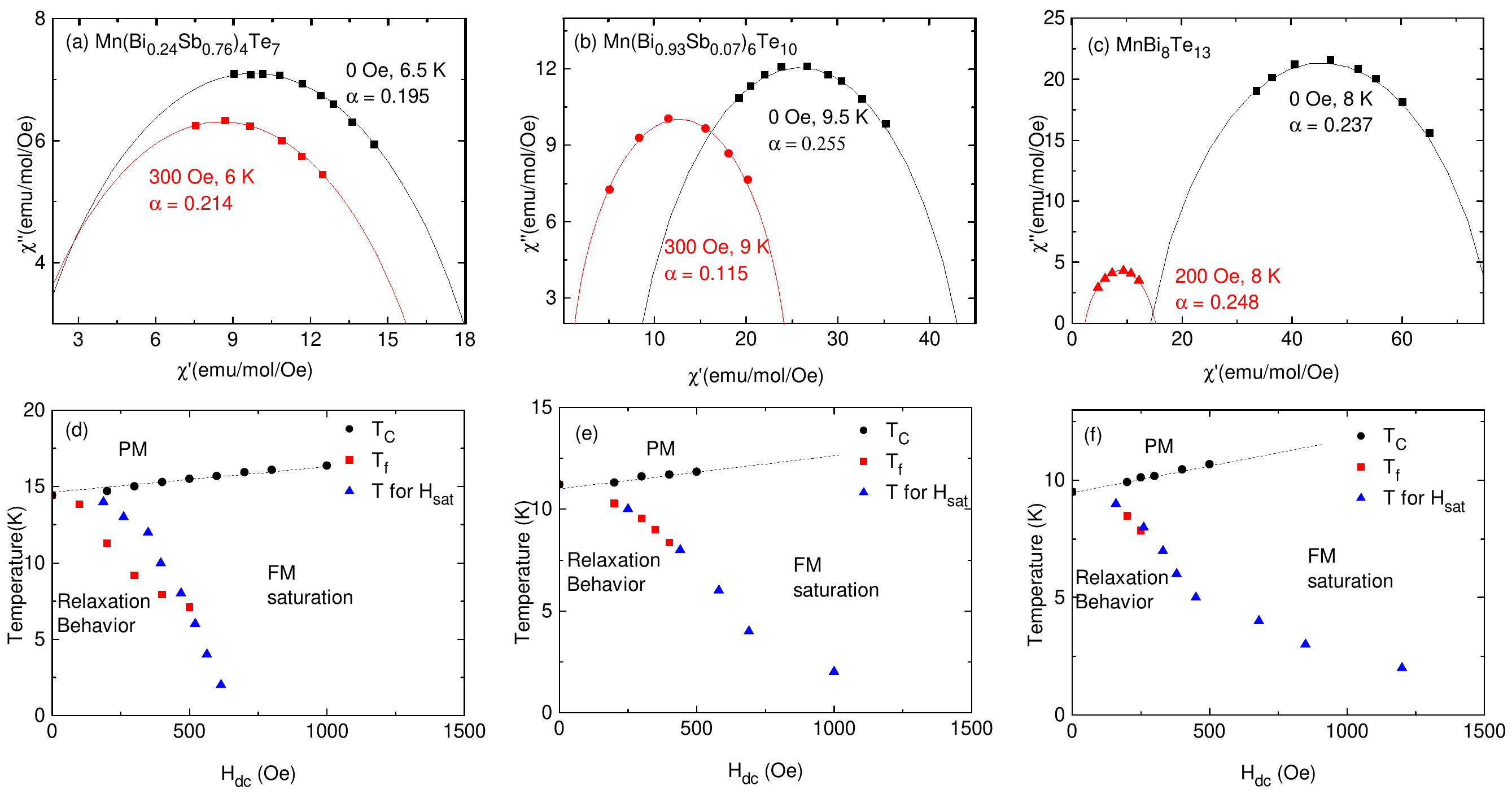} \caption{(a)-(c) Cole-Cole plots for Mn(Bi$_{0.24}$Sb$_{0.76}$)$_4$Te$_7$, Mn(Bi$_{0.93}$Sb$_{0.07}$)$_6$Te$_{10}$ and MnBi$_8$Te$_{13}$. $\chi''(\chi')$ data are taken from AC susceptibility at the selected temperature and DC field, across various frequencies. The fittings are done with Eqn. (4). (d)-(f) The DC field-temperature phase diagrams mapped via the AC susceptibility data. Also included are the saturation fields $H_{sat}$ obtained in the $M(H)$ measurements.}
\label{fig:ac-dc-pg}
\end{figure*}

To probe the non-steady state directly, we performed a DC scan to study the relaxation in MnBi$_8$Te$_{13}$ at constant temperatures over a long time. The measurement was taken after the sample was FC under 0.1 T to the target temperature and then the field was switched off. A clear relaxation of the overall magnetization can be observed. The time scale for the relaxation depends heavily on the temperature. As shown in Fig. 1 (g), it is on the order of hundred seconds at 4 K and tens of seconds at 3 K. However, anything faster than a few seconds becomes difficult to detect with this method, which motivates us to use the AC susceptibility to study the relaxation phenomena more comprehensively.

The temperature-dependent AC susceptibility of the three compounds measured under an AC field of $H_{ac}=10$ Oe with different frequencies are summarized in Figs. 2 (a)-(c). When $f=2$ Hz, $\chi'(T)$ has the similar shape and peak position to that of the ZFC $\chi(T)$ data, but it is narrower on the low temperature end. In addition, non-zero $\chi''(T)$ can be detected in all FM compounds. This confirms the presence of slow relaxation in the FM state, whose time scale is close to the oscillation period of the applied AC field. $\chi''(T)$ is peaked near 6 K, 9 K and 8 K for Mn(Bi$_{0.24}$Sb$_{0.76}$)$_4$Te$_7$, Mn(Bi$_{0.93}$Sb$_{0.07}$)$_6$Te$_{10}$ and MnBi$_8$Te$_{13}$ respectively, where $\chi'(T)$ has the steepest increase with temperature. With increasing AC frequency, the overall weight of the entire $\chi'(T)$ and the low-temperature-end of $\chi'(T)$ shift to higher temperatures, concurrent with the frequency-dependent peak shifts in $\chi''(T)$. Meanwhile, the peak position in $\chi'(T)$ signaling $T_C$ shows no frequency dependence and $\chi''(T)$ drops to zero at $T_C$, unlike that of a typical SG or SPM \cite{Mydosh1996,mydosh2015spin}. 

In attempt to single out the frequency dependent component, we then performed the AC susceptibility measurements at $H_{ac}=10$ Oe and $f=100$ Hz under various DC field bias ($H_{dc}$). The result is summarized in Fig. 2 (d)-(f). Remarkably, in all three compounds, an additional low-$T$ peak appears in $\chi'(T)$ under $H_{dc}$, resulting in the unusual ``double-peak" feature. As $H_{dc}$ increases, the original high-$T$ peak slightly moves to higher temperatures while the low-$T$ peak shifts to lower temperatures quickly. The magnitudes of these two peaks in $\chi'(T)$ decrease upon increasing $H_{dc}$ and the low-$T$ peak is completely suppressed with merely 600 Oe, 400 Oe, and 300 Oe in Mn(Bi$_{0.24}$Sb$_{0.76}$)$_4$Te$_7$, Mn(Bi$_{0.93}$Sb$_{0.07}$)$_6$Te$_{10}$ and MnBi$_8$Te$_{13}$, respectively. 

To better understand this rare AC magnetic behavior under $H_{dc}$, we measured the AC susceptibility with varying $f$ under a fixed $H_{dc}$. The data are shown in Figs. 2 (g)-(i). It turns out that the magnitude and the peak position of the high-$T$ peak are $f$-independent while the low-$T$ peak shows a strong frequency dependence, suggesting the low-$T$ peak is related to the freezing of some relaxation mechanism while the high-$T$ peak is associated with the long range ordering. Therefore, the temperature at the high-$T$ or low-$T$ peak maximum are labeled as $T_C$ or $T_f$, respectively. 

Next we focus on the low-$T$ peak. No matter the underlying causes for the slow relaxation, the overall
dynamics of relaxation is similar. Therefore, we can extract some characteristic parameters by analyzing the AC susceptibility data to differentiate various physical scenarios. From the frequency dependence of $T_f$, we can calculate the Mydosh parameter $K$, a measure of the relative peak shift per log frequency, by
\begin{equation}
    K=\frac{T_1-T_2}{T_{f0} (\log f_1-\log f_2)},
\end{equation}
where $T_1$, $T_2$ and $T_{f0}$ are taken as the $T_f$ at $f_1=2$ Hz, $f_2=200$ Hz and the DC limit (2 Hz), respectively. The obtained $K$ for each compound are 0.025, 0.016 and 0.020 respectively. The values are a few times greater than those in canonical SG with strong interaction and cooperative freezing such as Cu$_{1-x}$Mn$_x$ ($K=$0.005) \cite{Mydosh1996,mydosh2015spin}, but much smaller than those in the non-interacting SPM with gradual blocking such as $\alpha$-[Ho$_2$O$_3$(B$_2$O$_3$)] ($K=$0.28) \cite{gatteschi2006molecular,Mydosh1996,mydosh2015spin}, placing these materials in neither category.

Then we performed the Vogel-Fulcher fitting shown in the insets of Figs. 3 (g)-(i). The frequency dependence of $T_f$ can be fitted by the Vogel-Fulcher formula \begin{equation}
    \tau (T) =\tau_0\exp(E_a/k(T-T_0)).
\end{equation}
Here, $\tau$ is the most probable relaxation time for the system to overcome the energy barrier for spin reversal. At $T_f$, $\tau$ can be taken as $1/f$. $E_a$ is the thermal activation energy barrier for the spin reversal, $\tau_0$ is a time constant which is temperature-independent and usually larger than 10$^{-13}$s \cite{balanda2013ac}. $T_0$ is an empirical parameter in order to account for the deviation from a single-relaxation-time process due to the interaction between moments. Hence $T_0$ would be zero for an ideal non-interacting SPM. Yet fitting with $T_0$ = 0 would yield an unphysical value with $\tau_0 < 10^{-30}$s for all three samples. For the best fitting shown in the insets, $T_0$ is found to be around 8.1 K, 8.1 K and 7.1 K for the three samples. This again confirms that the relaxation behavior in these materials cannot be described by a single-relaxation-time process, but rather with a certain distribution of relaxation times. 

To quantify the spread of the relaxation time, we performed Cole-Cole fitting \cite{cole1941dispersion, balanda2013ac}. The generalized Debye formula describes the dynamic susceptibility as, 
\begin{equation}
   \chi_{ac}(\omega) \equiv \chi'-i\chi''=\chi_S+\frac{\chi_T-\chi_S}{1+(i\omega\tau)^{(1-\alpha)}}
\end{equation}
Here $\omega=2\pi f$, $\chi_S$ and $\chi_T$ are the high frequency and static constants, respectively, and $\alpha$ is an empirical parameter to account for the variation of the relaxation time around $\tau$ in the system. We can write down
\begin{multline}
    \chi''(\chi')=-\frac{\chi_T-\chi_{S}}{2\tan[(1-\alpha)\pi/2]}+
    \\ \sqrt{(\chi'-\chi_{S})(\chi_T-\chi')+    (\frac{\chi_T-\chi_{S}}{2\tan[(1-\alpha)\pi/2]})^2}
\end{multline}
$\alpha$ would be 0 for an ideal SPM with single-relaxation-time process, and around 0.9 for a SG system whose relaxation time has a broad variation. We extracted the $\chi''(\chi')$ data at selected temperatures from Figs. 2(a)-(c) and 3(g)-(i), and fitted the data using Eqn. (4) with $\alpha$, $\chi_S$ and $\chi_T$ as the fitting parameters. The results are shown in Figs. 3 (a)-(c). The obtained $\alpha$ are below 0.3, far from the expected value for SG ~0.9\cite{Mydosh1996}. The values are also consistent with the previous reports on Mn(Bi$_{0.7}$Sb$_{0.3}$)$_6$Te$_{10}$\cite{wu2020toward}. Therefore, similar relaxation process with $\alpha<0.3$ is not just limited to systems with large SL-SL separations but likely to be present among all members with FM coupling in this family.

So far we have shown that 1) the Mydosh-parameter estimation places these three compounds in neither SG nor ideal SPM type; 2) the Vogel-Fulcher fit suggests that the relaxation time has variations, excluding the ideal SPM scenario; 3) the fit of the Cole-Cole plot leads to a relatively moderate $\alpha<0.3$, excluding the SG scenario. This leaves us with two possible scenarios causing the relaxation behavior, the irreversible domain wall movement and cluster SG. To differentiate these two, we constructed the $T$-$H_{dc}$ phase diagram which mapped out $T_C$, $T_f$ and the corresponding $H_{dc}$ shown in Figs. 2(d)-(f). The obtained phase diagrams are shown in Fig. 3(d)-(f). Also included are the saturation fields obtained in $M(H)$ from Fig. 1 (b)(d)(f) at different temperatures. Interestingly, the $T_f$ line matches or stays below the data line obtained from $M(H)$ in the field regime where the low-$T$ peaks appear. This observation suggests the relaxation behavior only occurs before the magnetization saturates, that is, when there are irreversible domain wall motions\cite{topping2018ac}.

\subsection{Magneto-optical imaging of ferromagnetic domains}

\begin{figure}
\centering
\includegraphics[width=8.5cm]{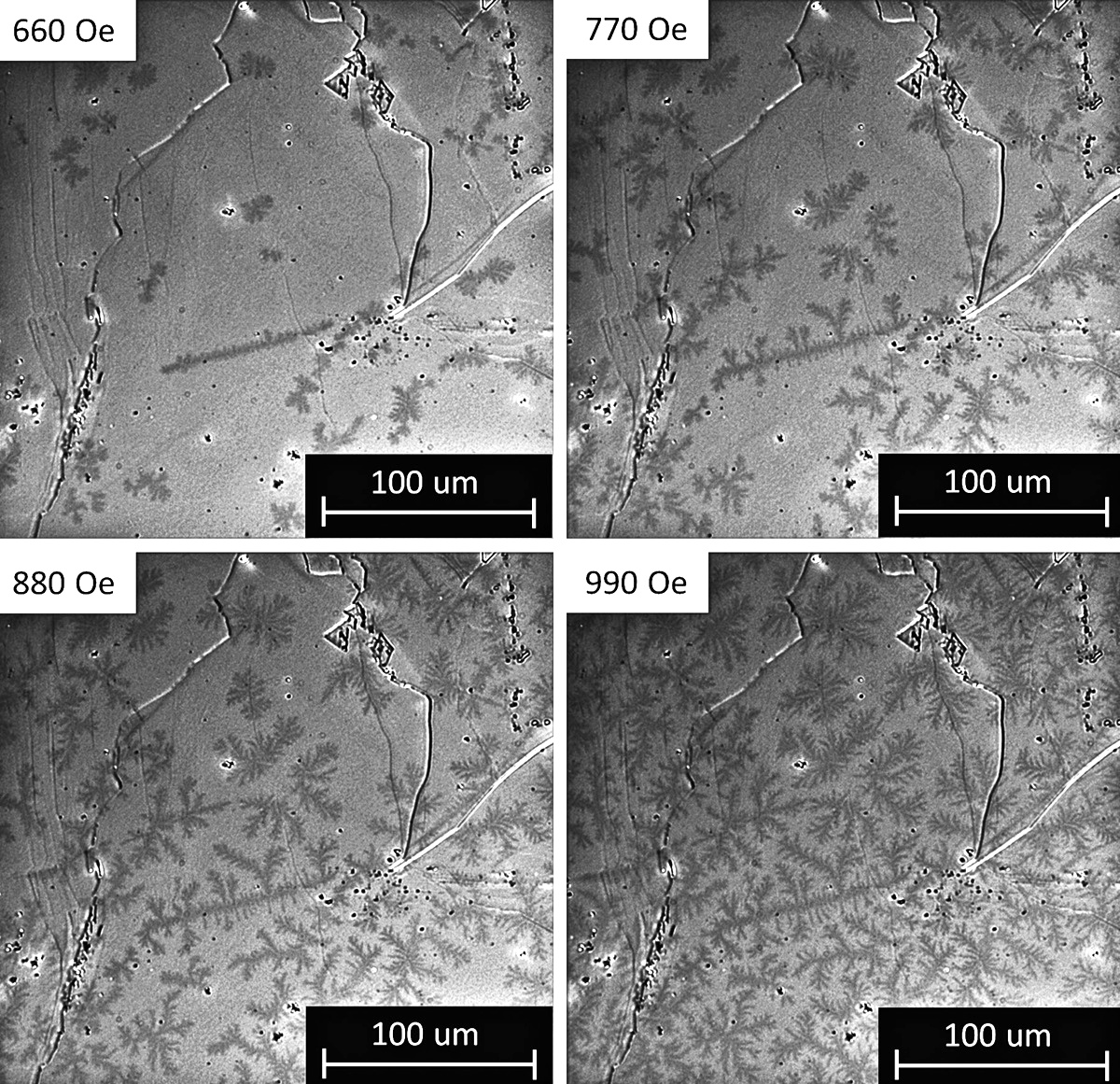}
\caption{Magneto-optical polar Kerr effect imaging of the ferromagnetic domains
in MnBi$_8$Te$_{13}$. After cooling in zero magnetic field to 5.2 K
( ZFC) a magnetic field of indicated amplitude
was applied. Up to about 600 Oe, no domains appear. After that dendritic
domains show up with a distinct one-dimensional grows along the dendrite
tips.}
\label{figLoop5K}
\end{figure}

\begin{figure}[tbh]
\centering
\includegraphics[width=8.5cm]{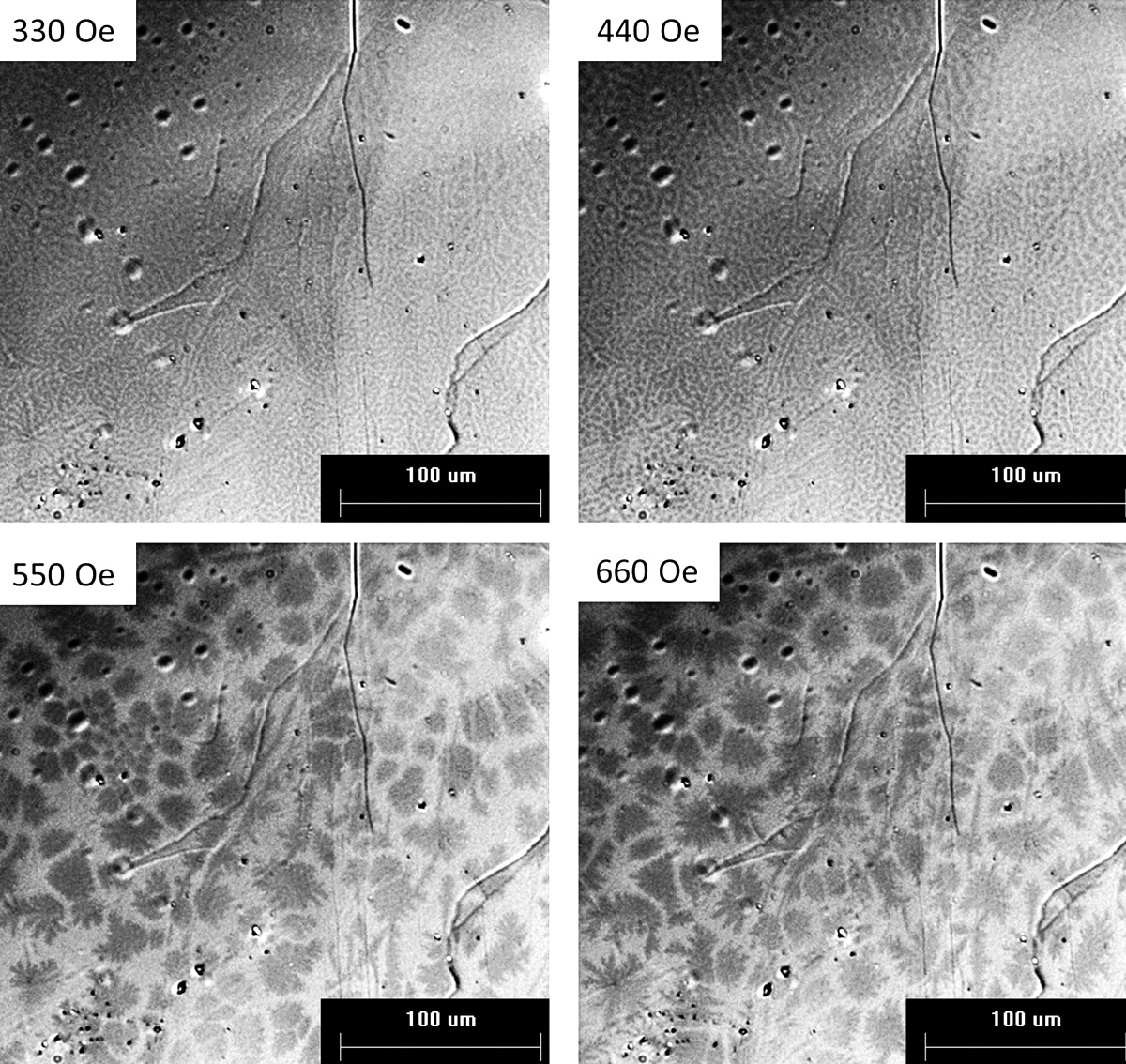} \caption{The remanent-state ferromagnetic domains in MnBi$_8$Te$_{13}$ crystals
at 5.2 K and H=0. The indicated magnetic field was applied and then
removed, mimicking the ``minor'' hysteresis loop to investigate
the irreversible coercive response. There is a significant change of
the domain structure around 500 Oe with higher field triggering the
appearance of sea-urchin domains. }
\label{figDomains5K}
\end{figure}

To directly visualize the domain shapes, magneto-optical imaging was performed. The FM nature of MnBi$_8$Te$_{13}$ is unambiguously established by
the direct observation of the FM domains shown in Figs. \ref{figLoop5K} and \ref{figDomains5K}. 

Figure \ref{figLoop5K} illustrates the appearance and evolution of the FM
domains when a DC field is applied after the sample was ZFC to 5.2 K, well below the FM transition. This
is a direct polarized light imaging with the contrast enhanced by setting
the gray-scale levels using the imaging processing software, but no structural
alterations were introduced and no local corrections were performed. With the increasing applied magnetic field, the first dendrite-looking domains appear at around 650 Oe and then the dendrites grow mostly in a one-dimensional fashion, similar to freezing ice.
Eventually, at higher magnetic field the entire area is filled with
the dark domain along the applied field. The dendritic
growth is magnetically very soft, because the domains propagate and
grow with almost no lateral displacement of the domain wall, spearheaded
by the domain tip. 

Note that a direct comparison with the $M(H)$ curve shown in Fig. 1(f) indicates a somewhat lower range of magnetic coercivity field compared to the imaging in Fig.~\ref{figLoop5K}. The difference is expected and is due to the difference in sample thickness. For imaging, very thin samples were used in search for the clean surface area, whereas magnetic measurements were performed on thicker samples. Initially, in a mono-domain state, the effective magnetic field on the edge is the sum of the applied field, $H_0$, and demagnetizing field, $H_{edge}=H_0-NM$, where $M$ is volume magnetization and $N$ is the effective demagnetizing factor \cite{ProzorovKogan2018}. For a FM sample, $M$ is positive and the effective edge field is reduced. Therefore, for a thinner sample where $N$ is larger than that of a thicker sample, the difference in the effective field can be quite substantial. In the future, for a more quantitative analysis, we will need to image and measure magnetization of the same sample. Here, the imaging is used as an unambiguous proof of long-range FM and it establishes a very soft nature of the magnetic domains to explain the peculiarities of the AC response.

Figure \ref{figDomains5K} further demonstrates the soft nature and
low coercivity of the FM state of MnBi$_8$Te$_{13}$ crystals. In
this experiment, after ZFC to $T=5.2$
K, the magnetic field shown in the legends was applied and then turned
off (reduced to zero). So, all images in Fig. \ref{figDomains5K} that are
taken at $H=0$ show the structure of the magnetic remanence. While in the ZFC imaging shown in Fig. \ref{figLoop5K}, a threshold field above 600 Oe is needed for the actual domains to appear, here we also see that the remanent-state domains do not appear up to about
300 Oe, probably indicating no domains or unsolvable
fine domain structure upon minor $M(H)$ loop when the field is reduced
to zero. The suggestion of a very fine structure finds indirect support in the images obtained after 330 Oe and 440 Oe were applied and
removed. We see a labyrinth-like structure, typical of very soft FM
material, such as Permalloy \citep{Hubert2008}. Only
above 500 Oe, the distinctly different, sea-urchin-looking structure,
appears and is stabilized at the higher fields with ``spines'' clearly
protruding off the body of the urchin-domains, suggesting small domain wall energy. Meanwhile, this pattern, when
the domain of one sign (of local magnetization) is embedded into the
continuous matrix of another dominant domain, is characteristic of
very soft magnetic response with small domain wall energy \citep{Hubert2008}.

Similar domain structures are reported for MnSb$_2$Te$_4$ \cite{ge2021direct}, so the soft FM is universal in all FM Mn-Bi-Te members.

\section{Discussion}

Based on our observations of 1) the relaxation behavior arises from the irreversible domain movement as indicated in Figs. 3(d)-(f), and 2) FM in this family is very soft and the domains are quite weakly pinned as revealed in Figs. 4 and 5, we can understand the distinct ``double-peak" AC response, including its $f$-dependence and $H_{dc}$-dependence shown in Fig. 2 as the following. Firstly, the high-$T$ $f$-independent peak arises from the long-range FM ordering, which should show no AC frequency dependence and monotonic increase $T_C$ with $H_dc$ in Fig 2 (d)-(i). Secondly, the low-$T$ $f$-dependent peak is due to the relaxation from the irreversible domain movements. According to Eqn. 2, the irreversibility increases with lowering temperatures. Thus, upon cooling, the domain regime crosses over from saturated single domain (reversible) to the emergence of domains, and then to the peak at $T_f$ where $\tau(T_f)=1/f$ and finally viscously slowing down and essentially freezing out with no response to the AC field at the lowest temperatures, resulting in the relaxation peak observed. 

This low-$T$ peak should be $f$-dependent because the viscous
force experienced by the domain walls is proportional to the instantaneous domain wall
velocity, thus to the frequency. Although the relaxation is from domain formations, rather than the scenario of SG or SPM, the overall
dynamics is similar to the SG, even more so, to a system of SPM nanoparticles where instead of magnetic clusters here we have very soft and practically isotropic magnetic domains.
Similar to the nanoparticles, domains are subject to a magnetic random
potential pinning landscape with a spread of energy barriers
for the thermally-activated relaxation, resulting in an effective barrier
that depends on the driving force. This mechanism leads to $f$-dependent
characteristic $T_f$ where the temperature
scales with the log of the frequency as shown in the inset in Figs. 2(g)-(i) and a moderate $\alpha$ between 0.1 to 0.3 in the Cole-Cole fitting as shown in Fig. 3(a)-(c)
\citep{prozorov1999,prozorov2004}.

The low-$T$ peak should also be $H_{dc}$ dependent. When $H_{dc}$ is increased, the single domain region expands to lower temperatures as shown in Figs. 1(b), (d) and (f), so the low-$T$ relaxation peak separates from the high-$T$ ordering peak. The signature of peak separation in $\chi'$ upon $H_{dc}$ was observed in the other soft FM FeCr$_2$S$_4$\cite{tsurkan2001ac}. However, it is more significant here since the FM here is softer. Such a small energy scale is a common feature in the FM members of MBT systems, likely due to the quasi-two-dimensionality and weak interlayer coupling.

Now let us understand the anomalous FC $\chi(T)$ and the ``bow-tie"-shaped hysteresis. Figure 5 reveals two types of domains, one is the fine-structured one and the other is the sea-urchin one. Although both types of domains are very soft, the sea-urchin one is less isotropic and thus more irreversible. Therefore, sea-urchin domains appear and dominate at lower temperatures/higher fields while the fine-structured ones emerge at higher temperatures/lower fields. This observation shed light on the non-trivial bow-tie shaped hysteresis observed. At lower fields, the fine-structured domains dominate with very small hysteresis while at higher fields, the emergence of the sea-urchin domains leads to larger hysteresis. Following the line, the unusual slope change in FC $\chi(T)$ thus likely separates the high-temperature fine-structured domain regime from the low-temperature sea-urchin domain regime where large bifurcation of ZFC and FC data appear. Future temperature-dependent magneto-optical imaging measurements will help verify this picture. 

The origin of the existence of two-types of domains may be related to the level of coupling between adjacent magnetic layers. At high temperatures or low fields, the interlayer coupling is weak comparing to the thermal fluctuations, so each magnetic layer forms individual domains, resulting in the fine-structured domains. When FC to low temperatures or under higher fields, the interlayer coupling wins over the thermal fluctuation, leading to the larger sea-urchin-shaped domains. Future measurements such as $\mu$SR are encouraged to illuminate this scenario.

\section{Conclusion}

Slow relaxation dynamics is observed in the DC and AC susceptibility measurements for all FM members in the MBT family. Such a phenomenon arises from the irreversible domain movement below the saturation field, in these soft ferromagnets with very weak pinning. The magneto-optics provides the direct evidence for the magnetic softness, manifesting as two types of isotropic remanent-state domains, a very fine-structured one and a much larger sea-urchin one. The former tends to dominate at lower fields, and the opposite for the latter, which may explain the anomalous bow-tie-shaped hysteresis loop and the slope change in the FC temperature-dependent susceptibility. Such knowledge of the domains will be essential for the ongoing pursuit of high temperature QAHE and other topological phenomena in the MBT device.

\section*{Acknowledgments}
NN and CH thank Prof. Arthur P. Ramirez for the insightful discussion as always. 
Single crystal growth, characterization and AC susceptibility measurements at UCLA were supported by the U.S. Department of Energy (DOE), Office of Science, Office of Basic Energy Sciences under Award Number DE-SC0021117. MOKE imaging at Ames lab and ISU was supported by the U.S. DOE, Office of Basic Energy Sciences, Materials Science and Engineering Division through the Ames Laboratory. The Ames Laboratory is operated for the U.S. Department of Energy by Iowa State University under Contract No. DE-AC02-07CH11358.

\medskip

\bibliographystyle{apsrev4-1}
\bibliography{SbMBT}
\end{document}